\documentclass[twocolumn,aps,showpacs,prb,10pt]{revtex4-1}
\usepackage{mathrsfs}
\usepackage{amsmath}
\usepackage{amsfonts}
\usepackage{amssymb}
\usepackage{amsthm}
\usepackage{graphicx}
\usepackage{graphics}

\begin{document}
\title{Domain Walls as Spin Wave Waveguides}
\author{X.S. Wang}
\author{X.R. Wang}
\email[Corresponding author:]{phxwan@ust.hk}
\affiliation{Physics Department, The Hong Kong University of
Science and Technology, Clear Water Bay, Kowloon, Hong Kong}
\affiliation{HKUST Shenzhen Research Institute, Shenzhen 518057, China}

\begin{abstract}
We numerically demonstrate that domain walls can be used as spin wave
waveguides. We show that gapless spin waves bounded inside a domain wall
can be guided by the domain wall. For Bloch walls, we further show that the
bound spin waves can pass through Bloch lines and corners without reflection.
This finding makes domain-wall-based spin wave devices possible.
\end{abstract}

\pacs{75.30.Ds, 75.60.Ch, 75.78.Fg, 85.70.Kh}
\maketitle
Magnonic spin current, or propagating spin wave, is a useful control
knob for spin manipulation. Compared to its electronic counterpart, it
has the advantages of low energy consumption and long spin coherence
length. \cite{YIG,Kajiwara} Unlike the electronic spintronic devices
that use ferromagnetic metals, magnonic devices allow the usage of
magnetic insulators which have very low damping relative to that of
magnetic metals. Controlled propagation of spin waves is an important
issue in magnonic devices. Analogous to wires conducting electric
currents, it is of great interest to have spin wave waveguides
that can guide spin waves to propagate from one place to another.
A good spin wave waveguide should be able to confine spin waves within a
narrow region, have a broad band, and be capable of guiding spin waves
to make a sharp turn. It is also important to keep the coherence of
the spin waves in the waveguide for phase sensitive magnonic devices.
\cite{phaseshift,swreview} Currently, spin waves are often generated
by microwave magnetic fields and guided by magnetic strips. \cite{YIG}
Spin waves generated in this way are mainly magnetostatic spin waves \cite
{magnetostatic} whose wavelengths are in the order of micrometers to millimeters,
much longer than the nanometer wavelength of exchange spin waves.
Thus, exchange spin waves are more attractive candidates for the
information carriers in high density magnonic devices. It is inevitable for
spin waves to make sharp turns during their propagation in real devices.
Although both experiments and theoretical studies have showen \cite{curveex,
curveth} that a spin wave can propagate along a properly designed curved
magnetic strip with the assistance of an external electric current, the general
applicability of the idea in devices is questionable because of sensitivity of
the device's performance to spin wave frequency and device geometry.
Furthermore, the required electric current will consume much energy.

On many occasions, spin waves are localized and propagate in special
restricted regions. For example, magnetostatic spin waves have surface
modes that exist and propagate unidirectionally along the interface of
magnetic and nonmagnetic medias. Magnetic domain walls (DWs), can interact
with spin waves. \cite{phaseshift,tatara,magnon,thermo,mstt,entropy,Wangkl}
Naturally, the idea of using DWs as spin wave waveguides has emerged.
\cite{waveguide,xiaojiang} It is already known that the gapped extended
exchange spin waves can pass through a DW without reflection
and with a phase shift \cite{phaseshift,magnon,entropy} while gapless
bound spin waves are localized inside the DW. \cite{magnon,entropy}
In a 1D DW, there is only one bound state of zero energy
that can be regarded as a Goldstone mode. \cite{tatara}
In 3D DWs, the bound modes with a gapless spectrum can propagate along
the DW and give rise to a larger DW entropy. \cite{entropy}
In this paper, we show that DWs can be used to guide the motion of bound
spin waves whose energies are lower than the gap of the extended modes.
For Bloch walls, the spin waves can pass through Bloch lines (BLs) and
corners without reflection. The spin waves have a phase shift when
passing through a BL. Thus, BLs can be used to manipulate the phase
of spin waves in such a waveguide.
A spin wave circuit board is designed according to these findings.

\begin{figure}[!htb]
\begin{center}
\includegraphics[width=8.5cm]{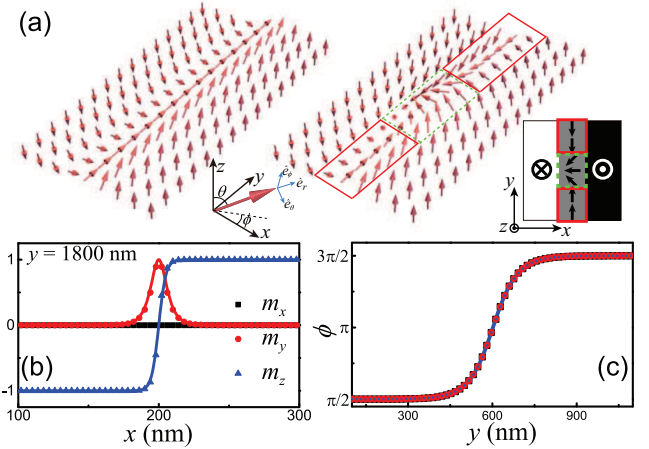}
\end{center}
\caption{(color online) (a) Schematic illustration of a Bloch DW in
perpendicularly magnetized films without (left) and with (right) a BL.
The coordinate system is shown in the middle. The red arrows denote
the magnetization directions. The lower right corner is the top view
of a Bloch wall with a BL. The BL (indicated by the green dashed box)
divides the Bloch wall into two segments with opposite polarities
(indicated by the red solid box). (b) The profile of a static Bloch DW
without the BL obtained from the \textsc{oommf} simulation (symbols)
at $y=1800$ nm and the Walker profile (solid lines).
(c) The $y$-dependence of $\phi$ at the DW center for a 400
nm$\times$ 3600 nm sample (black squares) and a 1200
nm$\times$1200 nm sample (red circles). Symbols are
\textsc{oommf} results and the solid line is the Walker profile
$\phi=\pi/2+2\tan^{-1}\left[e^{(y-Y)/\Delta_l}\right]$,
with $\Delta_l=54.4$ nm and $Y=1800$ nm.}%
\label{fig1}%
\end{figure}

We consider a 2D magnetic thin film in the $xy$-plane.
The magnetization dynamics are governed by the
Landau-Lifshitz-Gilbert (LLG) equation, \cite{LLG}
\begin{equation}
\frac{\partial\mathbf{m}}{\partial t}=-\gamma \mathbf{m}\times
\mathbf{H}_\mathrm{eff}
+\alpha \mathbf{m}\times \frac{\partial \mathbf{m}}{\partial t},
\end{equation}
where $\mathbf{m}$ is the unit vector of magnetization, $\gamma$ is
the gyromagnetic ratio and $\alpha$ is the Gilbert damping coefficient.
$\mathbf{H}_\mathrm{eff}$ is the effective field described as the variational
derivative of total energy density $E$ with respect to $\mathbf{m}$.
$\mathbf{H}_\mathrm{eff}=-\frac{\delta E}{\mu_0M_s\delta\mathbf{m}}$,
where $M_s$ is the saturation magnetization.
Due to the large magnetic anisotropy, perpendicularly magnetized films
have narrower DWs and better thermal stability. \cite{perp_record,perp}
Thus, we consider a single-layer perpendicular medium with a
uniaxial perpendicular (along the $z$-direction) anisotropy $K_u$.
The magnetostatic energy can be approximated by a thin-film shape
anisotropy energy $\mu_0M_s^2 m_z^2/2$. \cite{Hubert} Thus, the total
effective anisotropy is characterized by $K_z=K_u-\mu_0M_s^2/2>0$.
The energy density $E$ is then given by
\begin{equation}
E=A|\nabla \mathbf{m}|^2-K_zm_z^2,
\end{equation}
where $A$ is the exchange constant. Without dissipation,
the LLG equation becomes
\begin{equation}
\frac{\partial\mathbf{m}}{\partial t}=-\frac{2\gamma}{\mu_0M_s}\mathbf{m}
\times \left[A(\nabla^2m_x \hat{x}+\nabla^2m_y\hat{y})+K_z m_z\hat{z}\right],
\label{LLG2}
\end{equation}
where $\hat{x}$, $\hat{y}$ and $\hat{z}$ are unit vectors along the $x$-,
$y$- and $z$-axes. The static solution $\mathbf{m}_0$ of Eq.
\eqref{LLG2} for a down-up DW is the Walker profile,\cite{Walker}
$m_x=0$, $m_z=-\tanh\frac{x-X}{\Delta_w}$,
$m_y=\operatorname{sech}\frac{x-X}{\Delta_w}$,
where $X$ is the DW center position and $\Delta_w=\sqrt{A/K_z}$
is the DW width. A spherical coordinate system can be defined
with unit vectors $\hat{e}_r=\mathbf{m}_0$, $\hat{e}_\theta$,
and $\hat{e}_\phi$, where $\theta$ and $\phi$ are polar and
azimuthal angles of $\mathbf{m}_0$, as shown in Fig. \ref{fig1}(a).
To obtain the wave function of a spin wave with a well-defined frequency
$\omega$, we consider $\mathbf{m}$ fluctuating around $\mathbf{m}_0$
as, with a small amplitude of $|m_\theta|\ll 1$ and $|m_\phi|\ll 1$,
\begin{equation}
\mathbf{m}=\mathbf{m}_0+\left[m_\theta(x,y)\hat{e}_\theta+m_\phi(x,y)
\hat{e}_\phi\right]e^{-i\omega t}.
\end{equation}
Linearizing Eq. \eqref{LLG2} (to the first order in $m_\theta$ and $m_\phi$),
$\psi\equiv m_\theta-im_\phi$ satisfies the Schr\"{o}dinger equation,
\begin{equation}
\left(\frac{\mu_0M_s}{2\gamma K_z}\omega-1\right) \psi = \left(-\Delta_w^2
\nabla^2-2\operatorname{sech}^2\frac{x-X} {\Delta_w}\right)\psi.
\end{equation}
The equation has propagating solutions with extended wave functions
\begin{equation}
\psi_e=\left(-iq_x\Delta_w+\tanh\frac{x-X}{\Delta_w}\right)e^{i(q_xx+q_yy)},
\end{equation}
which is labelled by wavevector $\mathbf{q}_e=(q_x,q_y)$. Their dispersion
relations are $\omega_e=\frac{2\gamma}{\mu_0M_s}(Aq_e^2+K_z)$ with
a gap $\omega_g=\frac{2\gamma K_z}{\mu_0M_s}$ ($q_e=|\mathbf{q}_e|$).
The equation has also another set of solutions,
\begin{equation}
\psi_b=\operatorname{sech}\frac{x-X}{\Delta_w}e^{iq_by},
\label{bound}
\end{equation}
which, characterized by $q_b$, are bounded inside the DW and
propagate along the DW (the $y$-direction). The dispersion relation
is $\omega_b=\frac{2\gamma}{\mu_0M_s} (Aq_b^2)$ which is gapless.
When the frequency is lower than $\omega_g$, only bound spin waves
are allowed, and they must be confined to propagate along the DW.
Thus, the DW can serve as a waveguide for these bound spin waves.
\begin{figure}[!htb]
\begin{center}
\includegraphics[width=8.5cm]{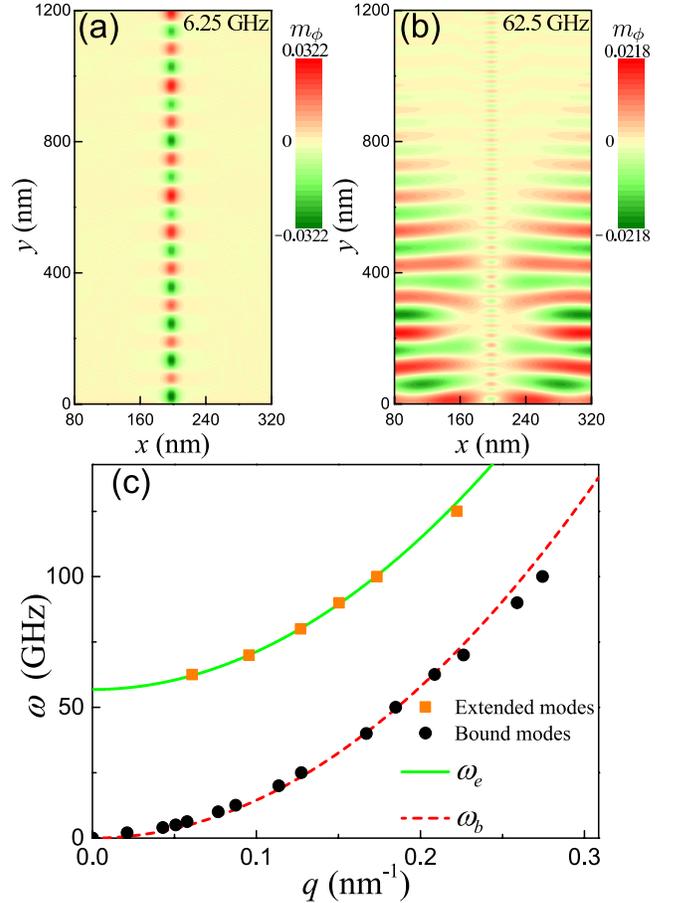}
\end{center}
\caption{(color online) (a) and (b): Density plots of $m_\phi$ at
$t=1.875$ ns for (a) $\omega_h=6.25$ GHz, (b) $\omega_h=62.5$ GHz.
The color codes $m_\phi$ values as indicated by the color bars.
(c) Dispersion relations of the extended modes and the bound modes.
Brown squares (Black circles) are numerical results for extended (bound)
states. Green solid line (red dashed line) are the corresponding
analytical formulas. }%
\label{fig2}%
\end{figure}

To verify the above theoretical proposal, we consider a thin magnetic
film of 400 nm$\times$3600 nm$\times$0.4 nm. The material parameters
are $A=15$ pJ/m, $K_u=0.8$ MJ$\cdot$m$^{-3}$, $M_s=580$ kA/m, and
$\gamma=28$ GHz/T which are widely used to mimic thin Co-film on Pt,
a well-known perpendicularly magnetized medium. \cite{Fert}
The magnetization dynamics of the film is simulated by using
\textsc{oommf} package \cite{oommf} with mesh size of 2 nm
$\times$2 nm$\times$0.4 nm. The damping $\alpha$ is set to 0.
The stable DW is Bloch-type, as illustrated in the left configuration of
Fig. \ref{fig1}(a) where the DW is centered at $x=200$ nm.
The $x$-dependence of three components of
$\mathbf{m}$ of a static DW at $y=1800$ nm are plotted in Fig. \ref{fig1}(b).
The symbols are from \textsc{oommf} simulations and the curves are the Walker
DW solution with $X=200$ nm and DW width $\Delta_w=\sqrt{A/K_z}=4.5$ nm.
The perfect agreement between the simulation results and the Walker
solution proves that the approximation of the shape anisotropy is good for
a Bloch wall which does not have bulk magnetic charge (on the contrary,
the Walker profile cannot describe a head-to-head N\'{e}el wall well
because the bulk magnetic charges modify the local anisotropy \cite{YHY}).
To generate spin waves, an oscillating magnetic field of
$\mathbf{h}=h\sin(\omega_h t)\hat{x}$ with $h=0.01$ T is applied in a
narrow region of $0\hbox{ nm}\leq y \leq 4$ nm.
Fig. \ref{fig2}(a) is a snapshot of spin wave distribution at 1.875 ns for
$\omega_h=6.25$ GHz which is below the frequency gap of $\omega_g=\frac
{2\gamma K_z}{\mu_0M_s}=56.8$ GHz according to  the early derivation.
Density plot of $m_\phi$ is coded by colors, varying from red for $m_\phi
=0.0322$ to green for $m_\phi=-0.0322$ with light yellow for $m_\phi=0$.
The spin wave exists only inside the DW as expected for a bound mode.
In contrast, for $\omega_h=62.5$ GHz which is above the gap, both extended
mode and bound mode are generated, as shown in Fig. \ref{fig2}(b) by the
density plot of $m_\phi$ coded by colors, varying from red for $m_\phi
=0.0218$ to green for $m_\phi=-0.0218$ with light yellow for $m_\phi=0$.
The extended spin wave propagates along the $y$-direction since the wave
fronts (the red and green strips) are parallel to the $x$-axis.
The wavelength of the spin wave in the DW is obviously shorter than that
of the extended spin wave of the same frequency (i.e. $q_b>q_e$), a
natural consequence of two different dispersion relations of spin waves.
Also, the bound spin wave propagates a longer distance than the extended
spin wave because the group velocity of the bound spin wave
$\frac{d\omega_b}{dq_b}=\frac{4\gamma A}{\mu_0M_s}q_b$ is faster than that
of the extended spin wave $\frac{d\omega_e}{dq_e}=
\frac{4\gamma A}{\mu_0M_s}q_e$ of the same frequency.
Fig. \ref{fig2}(c) shows the dispersion relations of the spin waves.
The numerical results (brown squares and black circles) are obtained
from the Fourier transforms of the $m_\phi$-$t$ curve at a fixed point
in the DW or in the domain for frequencies, and that of the $m_\phi$-$y$
curve at a fixed $x$ and $t$ for wavevectors.
The spin waves have the same frequencies as the applied field.
For both extended modes and bound modes, the numerical results are in
close agreement with the analytical formulas (green solid line and red
dashed line). The deviation between the simulation results and analytical
prediction is larger for larger $q_b$ and for the bound modes.
This is because the wavelength approaches the mesh size so that the
numerical results are less accurate.

\begin{figure}
\begin{center}
\includegraphics[width=8.5cm]{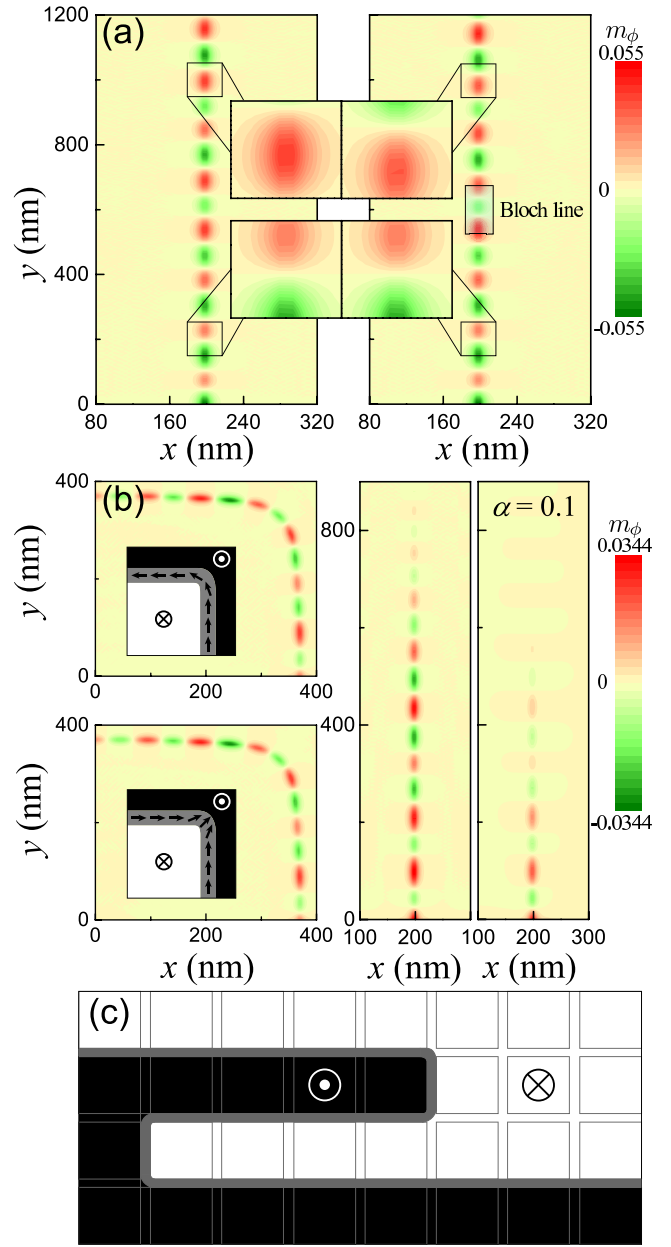}
\end{center}
\caption{(color online) (a) Density plot of $m_\phi$ at $t=1.875$ ns
for $\omega_h=4.0$ GHz and a Bloch wall without BL (left) and a Bloch
wall with a BL at $y=600$ nm (right). The inset shows the close-ups
of the regions indicated in the main figure for both samples.
(b) Density plots of $m_\phi$ at $t=0.625$ ns for $\omega_h=6.25$ GHz.
The left panel is for two kinds of curved DWs. The magnetization
configurations are shown in the insets. The middle panel is for a straight
Bloch wall. The right panel is for the same Bloch wall with $\alpha=0.1$.
(c) Schematic diagram of a complex circuit board.
The magnetization directions of the small squares are controllable.
Some squares are magnetized downward (white ones) or upward (black ones).
A curved DW (gray area) is between the two domains.}%
\label{fig3}%
\end{figure}

In reality, it is very difficult to prevent the BLs from nucleation
inside a transverse Bloch wall. \cite{Hubert,blochline,Ono}
A BL is the structure dividing two segments with opposite polarities in a
Bloch wall. It is important to know whether a Bloch wall with BLs can still
work as a spin wave waveguide. The magnetization texture of a Bloch wall
containing a BL is illustrated in the right half of Fig. \ref{fig1}(a).
In a thin film, a BL, called vertical Bloch line, is perpendicular to the
film with a well-defined width $\Delta_l$ that is independent of the
sample size. \cite{Hubert} This is verified in Fig. \ref{fig1}(c) by
plotting the $\phi$ angles of $\mathbf{m}$ at the DW center against $y$
for a 400 nm$\times$3600 nm$\times$0.4 nm sample and a 1200 nm$\times$1200
nm$\times$0.4 nm sample. If we put the BL centers in both samples at
$y=600$ nm, the two sets of data (black squares and red circles) overlap
each other. The $\phi$-$y$ plot can be well fitted by a Walker-like
solution $\phi=\pi/2+2\tan^{-1}\left[e^{(y-Y)/\Delta_l}\right]$, with
$\Delta_l=54.4$ nm and $Y=600$ nm. \cite{Hubert}
Fig. \ref{fig3}(a) shows the snapshots of $m_\phi$ at $t=1.875$ ns for
$\omega_h=4.0$ GHz. The left half is for a Bloch wall without BL, and
the right half is for a Bloch wall containing a BL at $y=600$ nm.
No reflection is observed in the frequency range of 2 GHz to 50 GHz in
our simulations. The inset shows close-ups of several regions indicated
in the main figure. Before the spin waves pass through the BL, the wave
profiles with and without a BL are identical, as shown in the lower row.
After the spin waves pass through the BL, the wave phase in the right
panel is lagged behind. The phase shift becomes smaller from $0.36\pi$
to about $0.07\pi$ as the frequency increases from 2 GHz to 50 GHz (to
calculate the phase shift, we track two crests emitted from the source
at the same time in the DW without and with a BL, and calculate the
difference in their positions when the later one reaches $y=1200$ nm).
Why the bound spin waves pass through a BL without reflection can be
understood by regarding a BL as a head-to-head 1D Walker-like wall within
a Bloch wall. \cite{DWinDW} Similar to that a Walker DW creates a
reflectionless potential well for extended spin waves, \cite{magnon} a BL
creates a reflectionless potential well for bound spin waves.
The details will be discussed in future publications.

We then investigate the propagation of the bound spin waves in a curved DW.
We use a 800 nm$\times$800 nm sample and apply a downward field
$\mathbf{H}=-0.1\hat{z}$ T in the region of $x\leq 360$ nm, $y\leq 360$ nm
and an upward field $\mathbf{H}=0.1\hat{z}$ T in the region of $x\geq 440$
nm, $y\geq 440$ nm to create a curved DW with a $90^\circ$ corner.
There are two types of such curved DWs with almost the same energy as shown
in the upper (lower) left inset in Fig. \ref{fig3}(b) where the DW turns
$90^\circ$ ($-90^\circ$) from the bottom end to the left end. The radius of
the corner curvature is about 70 nm. The snapshots of $m_\phi$ at 0.625 ns
for the two types of $90^\circ$-turning DWs are shown in the left panel of
Fig. \ref{fig3}(b). The frequency of the oscillating field is
$\omega_h=6.25$ GHz. The spin waves are clearly channeled in the curved DWs.
For a comparison, the snapshot of the spin wave at the same time for a
straight DW is shown in the middle panel.
By comparing the amplitudes of spin waves before and after passing through
the corner, no reflection or scattering at the corners can be identified.
This can also be seen by comparing the spin wave amplitudes in the curved
DW and the straight DW, which shows no significant change.
Our simulations show that for spin wave frequencies from 2 GHz to 50 GHz
(correspondingly, the wavelengths from 298 nm to 59.8 nm), the spin
waves pass through the corners without reflection.

Since the bounded spin waves can propagate in DWs without scattering even in
the presence of BLs and corners, one can use these properties to construct
complex spin wave circuit board, as shown in Fig. \ref{fig3}(c) as an example.
A magnetic thin film can be divided into arrays of small regions.
The magnetization of each region can be controlled by local magnetic fields or
exchange coupling with other materials, similar to the perpendicular magnetic
recording. \cite{perp_record} By programming the magnetization of each region,
complex DW structures can be created between up and down regions.
Spin waves can be externally generated and propagate along the DWs.
This domain-wall-based spin wave circuit board has the advantages of small
size and low loss. The spin wave phase can be tuned by controlling the
number and positions of BLs.

All the above results are still valid in the presence of damping despite
the spin waves decaying during their propagation.
The snapshot of $m_\phi$ for $\alpha=0.1$ with all other parameters
unchanged is shown in the right panel of Fig. \ref{fig3}(c). The decay of
the spin wave amplitude with the propagating distance can be observed.
From a device application viewpoint, it is better to use materials
with smaller damping. Also, one wants a larger perpendicular anisotropy
in order to have a broader frequency band and a thinner DW width.
However, if the anisotropy is too strong to make the DW width as thin
as the lattice constant, the Walker profile fails to describe this kind
of abrupt walls, and the spin wave may have different properties.
\cite{Yanpeng} Although we use Bloch walls to demonstrate the principle of
the concept, other kinds of transverse DWs such as in-plane head-to-head
walls or N\'{e}el walls can also be used as spin wave waveguides.
All the simulations shown above are performed for small $h$ in order to
eliminate nonlinear effects.
For larger $h$, the excited spin waves become irregular and nonlinear
effects become significant. For $\omega_h=6.25$ GHz, when $h$ is
as large as 0.2 T, harmonic spin waves can hardly be observed. However,
the BL can be driven to move along the same direction as the spin waves.
This phenomenon may have potential application in BL storage devices.
\cite{storage} Breaking of the chiral symmetry, such as
Dzyaloshinskii-Moriya interaction, can induce nonreciprocal spin waves
and leads to other interesting results. \cite{waveguide,xiaojiang}

In conclusion, we showed that the domain walls can be used as spin
wave waveguides when the frequencies of those spin waves are below the
frequency gap of those extended propagating spin waves. Inside the DW,
the spin wave can pass through the Bloch lines and corners without any
reflection although the phases will be shifted by the Bloch lines.
Complex spin wave circuits based on domain walls can be designed accordingly.

This work was supported by the National Natural Science Foundation of
China (Grant No. 11374249) as well as Hong Kong RGC Grants No.
163011151 and No. 605413.

\end{document}